\documentclass{aastex}
\usepackage{emulateapj5}
\usepackage{epsf}
\usepackage{onecolfloat}



\def\fsub{f_{\rm sub}}

\renewcommand\t{\times}

\begin{document}
\twocolumn[

\title{Lensing Optical Depths for Substructure and Isolated Dark Matter halos}

\author{Jacqueline Chen, Andrey V. Kravtsov \& Charles R. Keeton\altaffilmark{1}}
\affil{Department of Astronomy \& Astrophysics, Center for
Cosmological Physics, The University of Chicago,\\ 5640 S. Ellis Ave.,
Chicago, IL 60637}

\begin{abstract}
  
  Multiply-imaged quasar lenses can be used to constrain the
  substructure mass fraction in galaxy-sized dark matter halos via
  anomalous flux ratios in lensed images.  The flux ratios, however,
  can be affected by both the substructure in the lens halo and by
  isolated small-mass halos along the entire line-of-sight to the
  lensed source.  While lensing by dark matter clumps near the lens
  galaxy is more efficient than elsewhere, the cumulative effect of
  all objects along the line-of-sight could be significant.  Here we
  estimate the potential contribution of isolated clumps to the
  substructure lensing signal using a simple model motivated by
  cosmological simulations.  We find that the contribution of isolated
  clumps to the total lensing optical depth ranges from a few to tens
  percent, depending on assumptions and the particular configuration
  of the lens.  Therefore, although the contribution of isolated
  clumps to the lensing signal is not dominant, it should not be
  neglected in detailed analyses of substructure lensing.  For the
  currently favored $\Lambda$CDM model, the total calculated optical
  depth for lensing is high, $\tau\sim 0.2-20$ and could, therefore,
  naturally explain the high frequency of anomalous flux ratios in
  observed lenses.  The prediction, however, is highly sensitive to
  the spatial distribution of substructure halos in the innermost
  regions of the lens halo, which is still very uncertain.  Therefore,
  constraints on the properties of the substructure population or
  accurate cosmological constraints, such as the mass of the warm dark
  matter particle, are difficult if not impossible to derive at this
  point.

\end{abstract}

]
\altaffiltext{1}{Hubble Fellow}

\section{Introduction}

One of the generic predictions of the Cold Dark Matter (CDM) paradigm
is a ``clumpy'' distribution of matter, with a large number of
small-mass compact dark matter (DM) halos, both within virialized
regions of larger halos (substructure) and in the field.  At the
same time, the observed number of dwarf galaxy satellites in the Local
Group is more than an order of magnitude smaller than expected
\citep{klypin_etal99,moore_etal99}. This discrepancy has motivated
many theoretical studies of alternative models designed to reduce the
abundance of substructure
\citep[e.g.,][]{spergel_steinhardt00,hannestad_scherrer00,hu_etal00}
or to suppress star formation in small clumps via astrophysical
mechanisms, making them dark
\citep{bullock_etal00,benson_etal02,somerville02}.

If the CDM paradigm is correct, we expect $\sim 2-10\%$ of the mass of
a present-day galactic halo to be tied up in substructure \citep[e.g.,][]{klypin_etal99,ghigna_etal00,kravtsov_etal03}. The
absence of apparent optical counterparts would then indicate that the
small-mass DM clumps are dark. In the future, it may be possible to
detect substructure using different methods: for example,
indirectly via the effect of clumpy matter distributions on tidal tails of
satellite galaxies \citep{mayer_etal02} or directly via annihilation
or other observable signatures \citep[e.g.,][and references
therein]{bergstrom_etal99,iro_olinto02}.  At the current time,
however, gravitational lensing represents the best avenue to
constraining populations of DM clumps in galactic halos
\citep{mao_schneider98,metcalf_madau01,dalal_kochanek02,metcalf_zhao02}.
This is because substructure can modify the fluxes of lensed images
relative to those predicted by smooth lens models.  The existence of
such anomalous fluxes in many lens systems has been known for some
time \citep{mao_schneider98}. Recently, several such systems were
analyzed with the goal of constraining the properties of
substructure in lens halos.  \citet{dalal_kochanek02} carried out
a statistical study of seven radio lenses and found that the halo mass
fraction in substructure, $\fsub$, is approximately $\sim 2-5\%$
\citep[see][for an alternative interpretation]{evans_witt03}. This number
is broadly consistent with the typical substructure fractions of $\sim
0.02-0.1$ predicted by numerical simulations of CDM models
\citep{kravtsov_etal03}. Note, however, that predicted fractions are
for the total substructure populations within the virial radius, while
lensing observations are sensitive only to substructure in a cylinder of
radius $\sim 5-10$~kpc around the lens center.  In addition, the abundance of
clumps and their mass fraction can be expected to be depressed in the
central regions due to increased tidal disruption and merging.

\citet{dalal_kochanek02} and all other authors who have studied
substructure lensing have assumed that the lensing is caused only by
clumps in the halo of the lens galaxy. CDM models, however, predict
that large numbers of small-mass clumps also exist in the field around
galactic halos \citep[e.g.,][]{sheth_tormen99,klypin_etal99}. Although
the lensing efficiency of an individual field clump is expected to
be low, a typical path length from the source quasar may intersect the 
Einstein radii of many field clumps, resulting in a significant optical
depth relative to that of substructure lensing. Indeed, \citet{keeton03} recently showed that
the lensing efficiency for DM clumps, as a function of redshift, peaks at
the redshift of the lens but is fairly wide ($\Delta z\sim 0.3-0.5$)
and has tails that extend to the redshifts of the observer and of the
source.  Thus, the cumulative effect of all small-mass DM clumps
along the line-of-sight to the source could be significant.  In this
paper, we use a simple analytic model motivated by the results of
cosmological simulations to estimate and compare the optical depth for
lensing by substructure to that of the isolated 'field' halos. We
apply the model to two specific strong lens systems, B1422+231 and PG 1115+080, 
with anomalous flux ratios and present our conclusions for these systems. 

The paper is organized as follows. In section~\S\ref{sec:optdepth}, we
summarize the formalism of \citet{keeton03}, used to calculate the
lensing optical depth of individual DM clumps. We describe the model
for the spatial and mass distributions of satellite and isolated DM
halos in \S~\ref{sec:halos}. \S~\ref{sec:data} contains the main
details for the two observed lens systems that we analyze. Finally, we
present our results and conclusions in \S~\ref{sec:results} and
\ref{sec:conclusions}.

\section{Optical Depth}
\label{sec:optdepth}

The goal of this study is to compare the lensing efficiency of
substructure located within the lens halo to that of clumps along the
line-of-sight and in the immediate vicinity (but outside) the lens. We
do this by comparing the optical depth, a measure of the lensing
probability, of the two populations. The formalism for computing
the lensing optical depth for small DM clumps in a strong lens system was
recently derived by \citet{keeton03} and we refer the reader to this
paper for further details.  In this section we summarize the main
equations used in our calculations.

Since we are concerned with spatially small perturbations to a large
smooth gravitational lens, we assume that lensing cross sections can
be calculated by treating the lens as an external field. The
cross-section for lensing by clumps is
calculated assuming that their density distributions can be described
by singular isothermal spheres (SISs).  As the effect of clump
lensing can vary from negligible to very strong, we restrict ourselves to a
minimum threshold effect on image flux (i.e., magnification or
de-magnification).  To this
end, we estimate the cross section by taking into account all instances
where the
magnification or de-magnification 
by a clump is larger than a threshold value $\delta$.  We then calculate the total
optical depth for clump lensing by integrating the cross section over
the clump mass function in a given patch of sky and dividing by the 
area of the patch:
\begin{equation}
\tau(\delta; \kappa, \gamma) = \int dz D(z)^{2} \frac{dD}{dz} \int  dM \frac{dn}{dM} \sigma(\delta; \kappa, \gamma, \beta, M) \label{eq:tau},
\end{equation}
where $\kappa$ and $\gamma$ are the convergence and shear of the
image and $D(z)$ is the comoving distance.  $\beta$ parameterizes the distance ratios between the clump
and the halo, so that in the situation where the $z_1$ is the redshift of
the object closer to the observer and $z_2$ is the redshift
of the further object,
\begin{equation}
\beta = \frac{D_{12}D_{os}}{D_{o2}D_{1s}},
\end{equation}
where $D_{ij}=D(z_{i},z_{j})$ and $o$ and $s$ refer to the observer and source redshift, respectively.  

An estimate of the cross section for a general line-of-sight clump
is given by the effective area $A$ of the ``$\delta$-curve'', the
curve in the source plane where all images have
magnification $\mu = (1+\delta)\mu_0$, compared to the unperturbed 
magnification 
\begin{equation}
\mu_0 = [(1-\kappa)^2 - \gamma^2]^{-1}.
\end{equation}
This estimate gives the
exact cross section for configurations where the clump does not
change the number of lensed images, and a good approximation
(usually a lower limit) for configurations where the clump creates
additional faint, unresolved microimages. Specifically, the
estimated cross section $\sigma$ can be related to $A$ via
\begin{equation}
\sigma(\delta; \kappa, \gamma, \beta) = \left| \frac{{\rm det}(1-\beta \Gamma)}{{\rm det}(1-\Gamma)} \right| A(\delta; \kappa_{\rm eff}, \gamma_{\rm eff}), \label{cross_section}
\end{equation}
where the matrix $\Gamma$ is,
\begin{equation}
\Gamma = \left[
\begin{array}{cc}
\kappa + \gamma & 0 \\
0 & \kappa - \gamma 
\end{array}
\right].
\end{equation}
The area $A$ is a function of $\delta, \kappa$, and $\gamma$, where the convergence and shear are modulated by the difference in redshift
between the lens and the clump.  The effective values of the
convergence and shear are
\begin{equation}
\kappa_{\rm eff} = \frac{(1-\beta)[\kappa - \beta(\kappa^2-\gamma^2)]}{
(1-\beta\kappa)^2 - (\beta\gamma)^2}
\end{equation}
and
\begin{equation}
\gamma_{\rm eff} =\frac{(1-\beta)\gamma}{(1-\beta\kappa)^2 - (\beta\gamma)^2}.
\end{equation}

The functional form of $A$ depends on the global parity of the image
\citep[see][eqns. 11, 15, 17, and 18 for specific forms]{keeton03}.
For a negative parity image, the result is different for positive and
negative perturbations. Typically, the optical depth for the
$\delta<0$ case is much larger than the $\delta>0$ optical depth, so
clumps tend to make negative-parity images dimmer.

The dependence of the cross section on the mass of the clump is simple and is contained
within a factor of the Einstein radius squared, i.e., $\sigma \propto A \propto
b^{2}(M)$. For halos with a SIS density distribution, the Einstein radius is given by a simple expression:
\begin{equation}
b= 4 \pi \left( \frac{\sigma_{\rm rms}}{c} \right)^{2} \frac{D_{ls}}{D_{os}},
\label{eq:bsis}
\end{equation}
where $\sigma_{\rm rms}$ is the velocity dispersion of the halo, $c$
is the speed of light, and $D_{ls}$ and $D_{os}$ are the angular
diameter distances between the lens and the source and the observer
and the source, respectively. Due to the relative simplicity of the
SIS profile and corresponding expressions, we assume SIS
cross-sections in our analysis. Although SIS profiles do not describe
the density distributions of cosmological halos, this assumption is
reasonable for our purposes, since the density distribution of total
mass (baryons and DM) in the centers of galaxies is close to
isothermal. Also, at this point we are primarily interested in
evaluating the {\em relative} contributions of field clumps and
substructure rather than evaluating the absolute optical depth.  Here
we only note that the internal density distribution of substructure
halos is still uncertain and we are currently investigating it using
numerical simulations. The analysis presented in this paper can be easily
extended for other profiles.

\section{Populations of Dark Matter Halos}
\label{sec:halos}

We model the cosmological populations of dark matter halos by
separating halos into three distinct categories: 1) {\em substructure}
or halos located within the virial radius of the lens halo; 2) {\em
halos in the vicinity of the lens}: halos that are located outside the
virial radius but in the immediate vicinity of the lens; and 3) {\em
isolated} halos or halos that are located far from the lens. For the
latter halos, we assume a number density given by the average
cosmological halo mass function.  We do not take into account
substructure in other halos that projects onto the lens.  This is a
reasonable assumption the strong lenses analyzed in the literature
usually have no apparent bright host projecting near the lens.  At the
same time, in cases where a bright galaxy is seen near the lens, it is
included in the smooth mass model. Below we describe in detail how
each halo category was modeled.  Throughout the paper, we assume the
currently favored flat $\Lambda$CDM cosmology with the following
parameters: $h=0.65$, $\Omega_{M}$=0.35, $\Omega_{\Lambda}$=0.65, and
the power spectrum normalization of $\sigma_8$=0.92 (the {\em rms}
variance of mass distribution on the scale of $8h^{-1}{\rm Mpc}$).

\subsection{Substructure Halos}
\label{sec:substructure}

The substructure halo population is assumed to have a power-law mass function,
\begin{equation}
\frac{dn}{dm} \propto m^{-\alpha},
\label{eq:dndM}
\end{equation}
with $\alpha$ in the range of $1.7 - 1.9$, as measured in high-resolution cosmological
simulations \citep{ghigna_etal00}. The spatial distribution
of the substructure halos is assumed to follow that of dark matter:
i.e., their spherically averaged number density profile is assumed
to be described by a NFW profile \citep{navarro_etal97} outside of a core radius $r_{\rm c}$ and constant or zero inside the core radius:
\begin{equation}
\frac{dn}{dm} \propto
\left\{
\begin{array}{ll}
\left(r/r_{s}\right)^{-1}\left(1 + r/r_{s}\right)^{-2}, & r>r_{\rm c}\\
\rho(r_{\rm c}) ~{\rm or} ~0, &  r\leq r_{\rm c}.
\end{array}
\right.
\label{eq:nrsub}
\end{equation}
This model is based on the results of high-resolution cosmological simulations
\citep{colin_etal99,ghigna_etal00,kravtsov_etal03}. However, we should note
that predictions for the spatial distribution of substructure in
the innermost ($\lesssim 0.1 R_{\rm vir}$) regions of halos may still suffer
from ``overmerging'' (see discussion in \S~\ref{sec:conclusions}). As we will
show below, our results are very sensitive to the assumptions about 
the radial distribution of substructure halos and this aspect of the model is
one of its largest uncertainities. 

The characteristic inner radius is given by $r_{s}= R_{vir}/c_{vir}$,
where $R_{vir}$ is the virial radius of the lens halo corresponding to
an overdensity of $\Delta=340$ and $c_{vir}$ is the concentration
parameter as given by \citet{bullock_etal01}, as a function of mass
and redshift
\begin{equation}
c_{vir} = \frac{9}{1 + z_{l}} \left( \frac{M_{vir}}{M_{\ast}} \right)^{-0.13},
\label{eq:conc}
\end{equation}
where $M_{\ast}$ is defined as $\sigma(M_{\ast})=\delta_c=1.686$, and
$z_l$ is redshift of the lens halo.  The relation~(\ref{eq:dndM}) is
normalized by assuming that a certain fraction, $f_{\rm sub}$, of the
virial mass of the lens is associated with substructure and setting minimum
and maximum limits for substructure halo masses.  The normalization
constant, $K$, for the relation is dependent upon the mass limits, such that
\begin{equation}
K \propto  \left[ {M_{\rm max}}^{2 - \alpha} - {M_{\rm min}}^{2 - \alpha} \right].  
\end{equation}
The minimum mass, $M_{\rm min}$, is set to be very small and has
a negligible effect on our results for mass function slopes of $\alpha
< 2$ (it would, of course, be important for $\alpha=2$). For slopes that
approximate the mass functions of substructure in high-resolution
simulations, $\alpha\approx 1.7-1.9$, the normalization is sensitive
only to the maximum substructure mass, $M_{\rm max}$. For our fiducial
model (\S~\ref{sec:fiducial}) we assume that $M_{\rm max}$ is
equal to 0.01 of the host mass.  In cosmological simulations, $M_{\rm max}$ is 
typically found to be $\sim 0.01-0.05$.  Satellites of larger mass are
expected to merge with the host very quickly after accretion due to
dynamical friction.  In addition, in lens modelling
the most massive satellites are usually incorporated as part of the
smooth mass models.  We explore the sensitivity of the results to $M_{\rm
max}$ below (see Fig.~\ref{fig:taudep}).

As satellites accrete onto the lens galaxy, the matter in their outer
portions is tidally stripped. We approximate the tidal radius as the
radius at which the density of the clump equals the local density of the
lens halo.  The velocity dispersions are assumed to remain
unchanged.  Thus, the mass, $M$, for the substructure halos in
eq.~\ref{eq:dndM} is the mass within the tidal radius. This is taken into
account in calculating the Einstein radius and determining the mass
limits of integration. 

\begin{figure}[tb]
\centerline{\epsfxsize=3.2in \epsffile{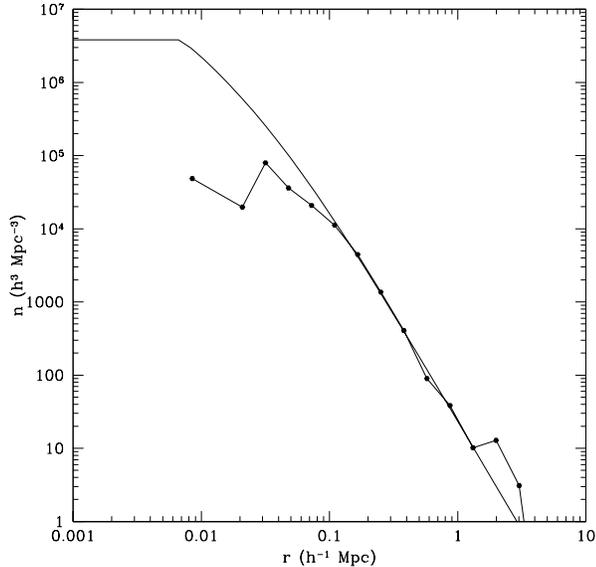}}
\caption{\footnotesize The number density profile of substructure
  halos predicted in our model (\S~\ref{sec:substructure}) compared to the
  number density profile measured in a high-resolution cosmological
  simulation of a galaxy-size DM halo.  The halo was simulated in the flat
  $\Lambda$CDM cosmology ($\Omega_m=0.3$, $h=0.7$, $\sigma_8=0.9$) and
  the profile measured at redshift $z=0.176$, typical for lens halos. At this
  epoch, the mass of the halo is $M_{180}=1.34\times 10^{12}h^{-1}{\ \rm
    M_{\odot}}$ and the most massive substructure halo has a mass of
  $M_{\rm max}=2.76\times 10^{10}h^{-1}{\ \rm M_{\odot}}$. We include
  only halos with $M_{\rm max}>M>M_{\rm min}=2.0\times 10^7h^{-1}{\ \rm
    M_{\odot}}$, which corresponds to the completeness limit of the
  halo catalogs. The figure shows that the agreement between our model
  and simulation result is good at large radii.  At small radii,
  however, the density of substructure halos is overestimated in the
  fiducial model; the simulation profile is much better approximated
  by a model with a larger value of $r_{\rm c}$. \label{fig:nrsim}}
\end{figure}

Figure~\ref{fig:nrsim} compares the number density profile of
substructure halos predicted in our model to the number density
profile measured in a high-resolution cosmological simulation of a
galaxy-sized DM halo \citep[one of the three halos analyzed
in][]{klypin_etal01}.  The halo was simulated in the flat $\Lambda$CDM
cosmology ($\Omega_m=0.3$, $h=0.7$, $\sigma_8=0.9$) and the profile
measured at $z=0.176$, typical for lens halos. At this epoch, the mass of
the halo is $M_{180}=1.34\times 10^{12}h^{-1}{\ \rm M_{\odot}}$ and
the most massive substructure halo has a mass of $M_{\rm max}=2.76\times
10^{10}h^{-1}{\ \rm M_{\odot}}$. We include only halos with $M>M_{\rm
min}=2.0\times 10^7h^{-1}{\ \rm M_{\odot}}$, which corresponds to the
completeness limit of the halo catalogs in the simulation. The mass
fraction in substructure halos within the virial radius for this halo is
$f_{\rm sub}=0.087$. 

The figure shows that the agreement between our model and the
simulation result is good at large radii.  At small radii, however,
the density of substructure halos is overestimated in the fiducial
model; the simulation profile is much better approximated by a model
with a larger value of $r_{\rm c}$.  However, the simulation may
actually underestimate the abundance of substructure in the inner
regions due to limited mass and force resolution (see also
\S~\ref{sec:conclusions}). We have chosen the fiducial model
conservatively, possibly overestimating the effects of
substructure. We estimate the effect of different assumptions about
the number density profile of substructure in \S~\ref{sec:results}. If
it turns out that substructure abundance has been overestimated our
conclusions about importance of isolated halos will be strengthened.

\subsection{Halos around the lens}

The number density profile of halos outside the virial radius but in 
the immediate vicinity of
the lens are accounted for in the calculation using the cross-correlation 
function
\begin{equation}
n_{m}(r) = \bar{n}_{m}[1+b_{M_{l}}b_{m}~\xi_{dm}(r)],
\label{eq:nrcrosscorr}
\end{equation}
where subscript $m$ indicates the number density of objects of mass
$m$, $b_{M_l}$ and $b_m$ are the bias for the lens halo and clump,
respectively, and $\bar{n}_{m}$ is the average cosmological density of
halos of mass $m$ (eq.~\ref{eq:st}).  We use expressions for the bias
and the mass function from \citet{sheth_tormen99} (see \S~\ref{sec:isolated} for details) and the fitting formula of
\citet{peacock_dodds96} to calculate the nonlinear dark matter correlation
function $\xi_{dm}(r)$.  Masses outside of the lens virial radius are
assumed to correspond to masses of overdensity $\Delta=178$ as in
\citet{sheth_tormen99}. The above expression is integrated for the mass limits assumed in the calculation (see \S~\ref{sec:fiducial}).  The total number density of dark matter halos as a function of radius
$n(r)$ is then the maximum of the number density from the lens substructure model (eqs.~\ref{eq:dndM} and ~\ref{eq:nrsub}) and the number density given by eq.~\ref{eq:nrcrosscorr} (see Fig.~\ref{fig:nr}).

We should note that the adopted model may overestimate the optical
depth of halos within and in the vicinity of {\em isolated galaxy-mass}
lenses. The equation~\ref{eq:nrcrosscorr} estimates the density
profile around an average halo, including halos in very dense regions
such as clusters.  Thus, the average profile $n_m(r)$ may be higher
than that of a relatively isolated halo.  Since many lenses are found
in groups and clusters \citep{keeton_etal00}, and the two specific
lenses we focus on here are located in galaxy groups, $n_m(r)$ should
provide a representative profile.

We find the optical depth by integrating the volume element for the
satellite halos -- both substructure halos and nearby halos -- over a
cylinder in Euclidean geometry, centered around the position of the
image, integrating through the halo out to a radius of $10h^{-1}\ \rm
Mpc$.  The optical depth equation, (\ref{eq:tau}), then becomes
\begin{equation}
\tau =  D_{ol}^2 \int dx \int dM \frac{dn}{dM}(M,x)~\sigma(M). \label{eq:tau_lt10}
\end{equation}

\subsection{Isolated Halos}
\label{sec:isolated}

For the population of isolated halos we assume a uniform spatial distribution
 with a mass function which
reasonably fits both the low mass and high mass ends of the mass function of halos
identified in numerical simulations of CDM models. Specifically, 
we use the analytic mass function of \citet{sheth_tormen99}
\begin{equation}
\frac{dn}{d {\rm~ln} M} = \frac{\rho}{M} \frac{d {\rm~ln}
\sigma^{-1}}{d {\rm~ln} M}f(\sigma),
\label{eq:st}
\end{equation}
where
\begin{equation}
f(\sigma) = A \sqrt \frac{2a}{\pi} \left[ 1 + \left(
\frac{\sigma^{2}}{a \delta^{2}_{c}} \right)^{p} \right]
\frac{\delta_{c}}{\sigma} {\rm exp} \left(- \frac{a \delta^{2}_{c}}{2
\sigma^{2}} \right),
\end{equation}
$A$=0.3222, $a$=0.707, $p$=0.3, and $\delta_{c}$=1.686.  We use the
fitting formula for the transfer function provided by \citet{eisenstein_hu99}
and a spherical top hat window function for the mass variance. 
As we noted above, all the halos are modeled as isothermal spheres,
where the masses of the clumps are defined as in
\citet{sheth_tormen99} to be the mass within the radius at which the
overdensity is $\Delta$=178.

To get the number of projected isolated clumps, we integrate over the
comoving volume from $z=0$ to the redshift of the source and over the
assumed mass limits for small-mass clumps:
\begin{equation}
\tau =  \int^{z_{s}}_{0} dz D(z)^{2} \frac{dD}{dz} \int dM \frac{dn}{dM}(M,z)~\sigma(M,z). \label{eq:tau_gt10}
\end{equation}

\subsection{Fiducial Model}
\label{sec:fiducial}
In our fiducial model, we set the magnification perturbation to
$\delta=0.2$, i.e., calculating the cross section for perturbations
greater than 20\%.  The mass limits for halos outside the virial
radius of the lenses are assumed to be $10^{-10} \times M_{vir}$ to
$0.1 \times M_{vir}$, where $M_{vir}$ denotes the virial mass of the
lens. Given that halos in a lens galaxy which pass close to the lens
center are typically stripped of $\sim 90\%$ of their mass by the
tidal field of the host after 1-2 orbits
\citep[e.g.,][]{klypin_etal99b,hayashi_etal02}, the upper limit on
substructure masses is assumed to be 0.1 of that for the isolated
clumps, i.e., $0.01 \times M_{vir}$.  The core radius of the
substructure number density profile is assumed to be $r_c=10$~kpc with 
a constant number density interior to it, the
slope of the mass function is set to $\alpha=1.8$, and the fraction of
mass in substructure is assumed to be $f_{\rm sub}=0.1$.  The value
of  $f_{\rm sub}$ is close to the upper end of the range found in
cosmological simulations and maximizes the normalization of the mass
function and the optical depth due to substructure. The sensitivity of
the results to $M_{\rm max}$, $f_{\rm sub}$, and $\alpha$ is shown in
Figure~\ref{fig:taudep} and will be discussed in
\S~\ref{sec:conclusions}.

\begin{table}[t]
\begin{center}
\caption{Lens Parameters \label{tab:data}}
\begin{tabular}{ccccc}
\tableline\tableline
\multicolumn{1}{c}{Lens/Image} &
\multicolumn{1}{c}{$\kappa$} &
\multicolumn{1}{c}{$\gamma$} &
\multicolumn{1}{c}{$\mu$} &
\multicolumn{1}{c}{image position} \\
 & & & & ($\arcsec$) \\
\tableline
B1422+231 & & & & \\
A                          & 0.384    &  0.476   &  6.57    &  1.014  \\
B                          & 0.471    &  0.634   &  -8.26   &  0.961  \\
C                          & 0.364    &  0.414   &  4.29    &  1.056  \\
D                          & 1.863    &  2.025   &  -0.30   &  0.284  \\
PG 1115+080 & & & & \\
$\rm A_{1}$                & 0.532    &  0.412   &  19.96   &  1.173  \\
$\rm A_{2}$                & 0.551    &  0.504   &  -19.10  &  1.120  \\
B                          & 0.663    &  0.644   &  -3.32   &  0.950  \\
C                          & 0.469    &  0.286   &  5.00    &  1.397  \\
\tableline
\end{tabular}
\end{center}
{\small Note -- $\kappa, ~\gamma, ~\mu$ are the convergence, shear, and parity of the images, respectively, from macromodels.  B1422+231 is modeled as a singular isothermal ellipsoid (SIE) plus an external shear \citep{keeton01}.  PG 1115+080 is modeled as a SIE plus an additional SIS representing the surrounding poor group of galaxies \citep{impey_etal98}. }
\end{table}

\begin{table*}[tb]
\tablenum{2}
\label{tab:tau}
\caption{Lensing optical depth due to different halo populations}
\begin{center}
\begin{tabular}{lcccc}
\tableline\tableline
\multicolumn{1}{l}{Lens/Image} & 
\multicolumn{1}{c}{$\tau_{\rm sub}$} &
\multicolumn{1}{c}{$\tau_{\rm <10} - \tau_{\rm sub}$} &
\multicolumn{1}{c}{$\tau_{\rm >10}$} &
\multicolumn{1}{c}{$\tau_{\rm >10}/\tau_{\rm <10}$} \\
\tableline
B1422+231 & & & \\ 
A                       &1.8            &$1.9\t10^{-3}$&$4.5\t10^{-2}$ &$2.5\t10^{-2}$ \\
B ($\delta < 0$)        &2.8            &$2.8\t10^{-3}$&$2.1\t10^{-2}$ &$7.6\t10^{-3}$ \\
~~~($\delta > 0$)       &$6.5\t10^{-3}$ &$6.7\t10^{-6}$&$4.2\t10^{-3}$ &0.65           \\
C                       &0.76           &$8.2\t10^{-4}$&$2.8\t10^{-2}$ &$3.7\t10^{-2}$ \\
D ($\delta < 0$)        &$1.1\t10^{-3}$ &$7.6\t10^{-7}$&$3.4\t10^{-7}$ &$3.2\t10^{-4}$ \\
~~  ($\delta > 0$)      &$2.9\t10^{-2}$ &$2.1\t10^{-5}$&$1.0\t10^{-2}$ &0.35           \\
PG 1115+080 & & & \\                                                       
$\rm A_{1}$             &15.            &$1.0\t10^{-2}$&$8.3\t10^{-2}$ &$5.6\t10^{-3}$ \\
$\rm A_{2}$ ($\delta<0$)&14.            &$1.0\t10^{-2}$&$5.6\t10^{-2}$ &$4.0\t10^{-3}$ \\
~~~~ ($\delta > 0$)     &$1.1\t10^{-2}$ &$7.6\t10^{-6}$&$6.6\t10^{-4}$ &$6.2\t10^{-2}$ \\
B ($\delta < 0$)        &0.41           &$2.7\t10^{-4}$&$9.3\t10^{-4}$ &$2.3\t10^{-3}$ \\
~~ ($\delta > 0$)       &$2.3\t10^{-2}$ &$1.5\t10^{-5}$&$1.5\t10^{-3}$ &$6.5\t10^{-2}$ \\
C                       &0.88           &$7.0\t10^{-4}$&$9.6\t10^{-3}$ &$1.1\t10^{-2}$ \\
\tableline
\end{tabular}
\end{center}
\end{table*}

\section{Lens Parameters}
\label{sec:data}

We estimate lensing cross-sections for two specific cases
of quadruple-image gravitational lenses in which flux anomalies have
been detected: B1422+231 and PG 1115+080.  The source B1422+231 is a
radio-loud quasar at $z$=3.62 lensed by an early-type galaxy in a poor
group of galaxies at $z$=0.34
\citep{patnaik_etal92,kundic_etal97b,tonry98}.  Images A and C are
bright positive-parity images, parity corresponding to the sign of the
magnification, while image B is a bright negative-parity image and D
is a faint negative-parity image \citep{patnaik_etal99}.
\citet{keeton01} and \citet{bradac_etal02} recently argued that a
clump projecting in front of the image A can fit the lens data,
concluding that mass of the perturber should be of order $\sim
10^{5-6} M_{\sun}$ if it is a point mass, or $\sim 10^{6-7} M_{\sun}$
if the mass distribution is extended (i.e., a halo) with a SIS
density distribution.

PG 1115+080 is a radio-quiet quasar at $z$=1.72 lensed by an
early-type galaxy in a poor group of galaxies at $z$=0.31
\citep{weymann_etal80,kundic_etal97a,tonry98}.  Images $\rm A_{1}$ and
C are positive-parity images, while $\rm A_{2}$ and D are
negative-parity images \citep{impey_etal98}.  Smooth lens models for
PG 1115+080 are able to fit all but the $\rm A_{1}/A_{2}$ flux ratio.
Table~\ref{tab:data} summarizes the properties of the observed images
in these studies relevant to our study.  The concentration parameter
for B1422+231 is 7.9, while that of PG 1115+080 is 7.3.  The
concentration parameters are estimated using eq. \ref{eq:conc} for the
virial mass of the lens and its redshift.  The mass, in turn, 
is derived from the lens model. Namely, we use the estimate of the
Einstein radius from the lens velocity dispersion and invert
eq. \ref{eq:bsis}. The virial mass is then calculated assuming an 
isothermal profile as the mass within the radius corresponding to
overdensity of $\Delta$=340.  The estimates of the
virial mass and the Einstein radius of B1422+231 is $5.1 \times
10^{12} M_{\sun}$ and $0.764\arcsec$
\citep{keeton01}, while those of PG 1115+080 are $1.1 \times
10^{13}M_{\sun}$ and $1.147\arcsec$ \citep{impey_etal98}.

\section{Results}
\label{sec:results}

Using the model described in the previous sections, we calculate and
compare the lensing optical depths due to the substructure halos,
$\tau_{\rm sub}$; due to all halos within $r<10h^{-1}$ Mpc of the
lens, $\tau_{<10}$ (this optical depth includes both substructure and
halos in the vicinity of the lens); and due to all 'field' halos along
the line of sight from the source to the observer but excluding halos
within $10h^{-1}{\rm\ Mpc}$ of the lens, $\tau_{>10}$.

Table~\ref{tab:tau} compares $\tau_{\rm sub}$, $\tau_{<10}$, and
$\tau_{>10}$ for our fiducial model. The table shows that in the
fiducial case the substructure halos provide a dominant contribution
to the lensing optical depth for all images. The contribution of
nearby clumps outside the virial radius of the lens (i.e.,
$\tau_{<10}-\tau_{\rm sub}$) is negligible.  The optical depth due to
isolated clumps along the line of sight, $\tau_{>10}$, is generally
small compared to the substructure optical depth (a few percent) for
all positive parity images.  In negative parity images, the relative
effect for magnification is much larger. Magnification, however, is a
much smaller effect than de-magnification in these images.  Thus, the
contribution of isolated halos is relatively small. It is, however,
not negligible and should be included in detailed analyses of
substructure lensing.

The results of the full numerical model can be recovered using
simple analytic estimates.  
The substructure optical depth can be estimated as (c.f. eq. \ref{eq:tau_lt10})
\begin{equation}
\tau_{\rm sub} \approx D_{ol}^{2} \Delta x \left[\frac{dn}{dM}(M_{\rm eff}) \right]\Delta M~ \sigma(M_{\rm eff}),  \label{eq:est_tausub}
\end{equation} 
where $\Delta M \approx M_{\rm eff}$.  Given that the substructure
optical depth ought to be of order unity, we can use this equation to
estimate the effective mass.  For B1422+231, $M_{\rm eff} \sim 10^{8}
M_{\odot}$, where $M_{\rm eff}$ is a tidally truncated mass.  The
corresponding virial mass of isolated halos would then be $M^{\rm
iso}_{\rm eff} \sim 10^{9} M_{\odot}$.  We can thus estimate the
relative importance of the line-of-sight optical depth
(c.f. eq. \ref{eq:tau_gt10}):
\begin{equation}
\tau_{\rm >10} \approx \Delta D D_{ol}^{2} \left[\frac{dn}{dM}(M^{\rm
iso}_{\rm eff},z_{l})\right]\Delta M~\sigma(M^{\rm iso}_{\rm eff},z_{l}),
\label{eq:est_taugt10}
\end{equation}
where for $\Delta z \sim 0.3 - 0.5$ (the wings of the lensing cross section), $\Delta D \sim D_{ol}$.  Rewriting the equation explicitly in term of mass,
\begin{equation}
\tau_{\rm >10} \approx 1.6 \times 10^{-25} D_{ol}^{3} \left[\frac{dn}{dM}(M_{\rm eff},z_{l})\right]M_{\rm eff}^{7/3}.
\end{equation}
For B1422+231 and the given effective mass of $M_{\rm eff} \sim 10^{9}
M_{\odot}$, $D_{ol} \sim 10^3$ Mpc and $\frac{dn}{dM}(M_{\rm eff})
\sim 10^{-9}$, $\tau_{\rm >10} \sim 10^{-4}$.  The optical depth is
$\tau\sim M_{\rm eff}^{1/2}$, so for $M_{\rm eff} =10^{11}
M_{\odot}$ the optical depth is $\tau\approx 10^{-3}$, roughly
consistent with our numerical results. The total optical depth is a
sum of the contributions by halos of different masses. This exercise
shows, however, that most of the signal for the model we adopted is
due to the high-mass end of the halo mass function.

\begin{figure}[tb]
\centerline{\epsfxsize=3.2in \epsffile{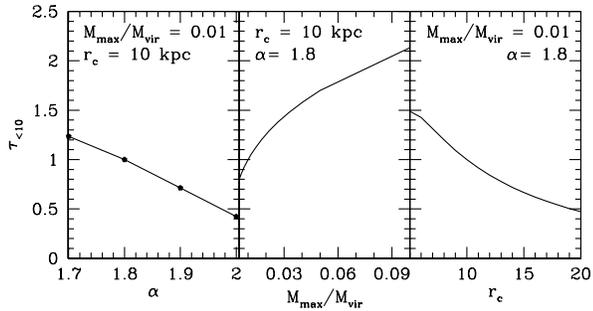}}
\caption{\footnotesize The effect of varying $\alpha$, the upper mass limit $M_{\rm max}$, and the core radius, $r_{c}$, on the optical depth of satellites is shown here from left to right for B1422+231, image A, where the optical depth is normalized to $\tau$=1 for the fiducial values. In each
case, all other parameters of the model are kept fixed at their fiducial 
values. \label{fig:taudep}}
\end{figure}

\begin{figure}[tb]
\centerline{ \epsfxsize=3.2truein\epsffile{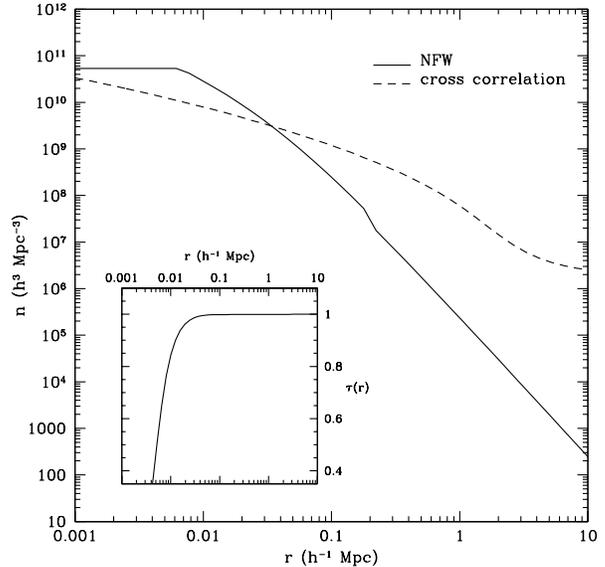} }
\caption{\footnotesize The number density profile for the image A of lens 
  B1422+231. The two lines show the profile for substructure halos
  (eq.~\ref{eq:nrsub}, solid line) with fiducial model
  parameters and the profile for the halos in the vicinity of the lens
  (eq.~\ref{eq:nrcrosscorr}, dashed line). The break in the profile of the substructure
  halos at $\approx 200h^{-1}{\ \rm kpc}$ (the virial radius) is due
  to change of mass from tidally truncated to virial (for halos
  outside the virial radius of the lens) in our model.  The
  corresponding optical depth profile is inset.}
\label{fig:nr}
\end{figure}

The relative contribution
of isolated halos could be larger, if the fiducial model overestimates
the abundance of substructure halos. Below we discuss the
sensitivity of the optical depth calculations to our model parameters.  
The optical depth $\tau_{\rm sub}$ depends on the assumed slope of the
halo mass function, $\alpha$; the mass limits, $M_{\rm min}$ and
$M_{\rm max}$; the mass fraction in substructure, $f_{\rm sub}$; and
the assumed shape of the radial number density profile.  The lensing
cross section $\sigma$ is proportional to the square of the Einstein
radius, which, for SIS halos, implies
\begin{equation}
\sigma \propto m^{4/3}.
\end{equation}  
Since the mass function of small-mass DM clumps is approximated by a 
power-law $n \propto m^{-\alpha}$, we have
\begin{equation}
\frac{d\tau}{dm} \propto m^{4/3 - \alpha}, ~\tau \propto  \left[ {M_{\rm max}}^{7/3 - \alpha} - {M_{\rm min}}^{7/3 - \alpha} \right].  
\end{equation}
For $\alpha=1.7-2$, $7/3 - \alpha$ is in the range $2/3-1/3$, and the
value of $M_{\rm min}$ is unimportant.  For the fiducial case, $\alpha=1.8$, $\tau_{\rm
  sub} \propto M_{\rm max}^{8/15}$.  The dependence on $f_{\rm sub}$
is simple: $\tau_{\rm sub}\propto f_{\rm sub}$.  

The constraints on substructure masses for B1422+231 suggest that the fluxes of this lens can only be affected by masses, $M \ge 10^{6} M_{\odot}$ \citep{keeton01,bradac_etal02}.  Thus, masses below this value but above the minimum mass, should have little effect on the optical depth.  The substructure optical depth for image A of the system for masses between $M_{\rm min}$ and $10^{6} M_{\odot}$ is $\tau_{\rm sub} = 0.005$, or 0.3\% of the total optical depth.  The optical depth for masses between $M_{\rm min}$ and $10^{7} M_{\odot}$ is $\tau_{\rm sub} = 0.02$, or 1\% of the total, calculated optical depth.  

The fiducial value of $f_{\rm
  sub}=0.1$ is typical for CDM halos. Although
$f_{\rm sub}$ varies from halo to halo, variations of more than a
factor of $\sim 2-3$ from the fiducial model are expected to be rare
\citep{kravtsov_etal03}.  Nevertheless, in the case when $f_{\rm sub}$ and $M_{\rm
  max}$ are at the lower end of the range expected for CDM halos the
optical depth can be reduced by a factor of $\sim 2$.  Note that these
parameters are not independent. For example, for $\alpha<2$ lower
$M_{\rm max}$ will correspond to lower $f_{\rm sub}$.

Figure \ref{fig:taudep} shows the effect of varying $\alpha$, $M_{\rm
max}$, and the core radius, $r_{c}$, on the optical depth due to
nearby and substructure halos relative to the fiducial model.  The
figure shows that varying the mass function slope $\alpha$ has
relatively small effect: less than a factor of 2 for a realistic range of
values.  The variation of $M_{\rm max}$ (for $\alpha=1.8$) results in
only a factor of $\sim 2-3$ change in optical depth.  In the case
where we model the substructure number density profile as described in
\S~\ref{sec:substructure}, our only variable parameter is $r_{c}$,
which also varies the optical depth by a factor of $\sim 2-3$.  Thus,
our results are not expected to change drastically due to variations
of $\alpha$, $r_{c}$, and $M_{\rm max}$. The optical depth profile,
however, is also affected, which we discuss below, and the shape of
the radial number density profile is not well-constrained in
cosmological simulations.

Figure \ref{fig:nr} shows the number density profile of clumps inside
and in the immediate vicinity of the halo of lens B1422+231; the inset
shows the cumulative optical depth profile  for image A,
which is likely to have substructure affecting its flux. The profile 
is normalized to unity and 
does not include the effect of isolated halos outside $10h^{-1}\ {\rm Mpc}$.
The figure shows that most of the optical depth is contributed by 
regions of the highest number density. In this particular 
case, the clumps outside $\sim$10\% of the virial radius contribute 
$\lesssim 2\%$ of the optical depth. Clearly, the substructure optical 
depth is very sensitive to the distribution of DM clumps in the innermost
regions of the lens halo. In our model the inner number density of substructure
clumps is determined by the core radius $r_{\rm c}$. 

\begin{figure}[tb]
\centerline{\epsfxsize=3.2in \epsffile{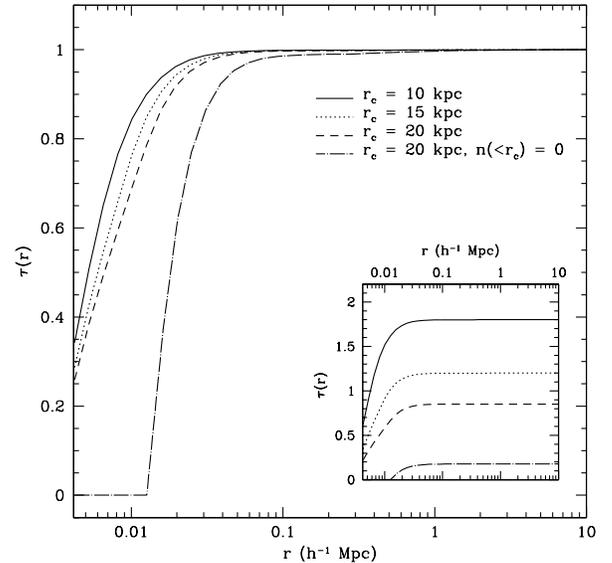}}
\caption{\footnotesize The optical depth profile for image A of lens 
  B1422+231 normalized to unity, for different values of $r_{\rm c}$. The solid
  line is for the fiducial value of $r_{\rm c}$=10 kpc, the dashed
  lines are $r_{\rm c}$=15 kpc, $r_{\rm c}$=20 kpc, and $n(r)$=0 when
  $r <$ 20 kpc, respectively.  The unnormalized profiles are
  inset. The contribution of isolated line-of-sight halos are not included
  in the optical depth shown here.\label{fig:nrrc}}
\end{figure}

Figure~\ref{fig:nrrc} shows the optical depth profiles for different
values of $r_{\rm c}$. The main panel shows the optical depth
$\tau(r)$ normalized to unity. Here again we see that halos at
$r\lesssim r_{\rm c}$ contribute most of the signal.  The inset panel
shows the corresponding unnormalized profiles $\tau(r)$ and
demonstrates that the total optical depth due to substructure is quite
sensitive to the inner distribution of clumps.  The lowest curve
corresponds to the case where no clumps are present
within the central 20~kpc. In this case the optical depth is a factor of
ten smaller than the optical depth for the fiducial model, which would
increase the relative contribution of isolated halos to $\sim$20-30\%
(c.f. Table~\ref{tab:tau}).

\section{Discussion and conclusions}
\label{sec:conclusions}

Substructure in galaxy-sized lens halos, if present, may alter the
flux ratios of gravitationally lensed images. Studies of multiply-imaged 
quasar lenses are therefore a unique probe into the small-scale
matter distribution in DM halos and may prove to be one of the most
powerful tests of the CDM paradigm. Indeed, two of the most glaring
problems for the CDM models are the density distribution in galaxies
(predicted halos are too dense and possibly too cuspy) and clusters
and the overabundance of small-mass DM clumps. Both problems are a
manifestation of the relatively high amplitude of the CDM power spectrum
on small scales.  If DM clumps can be proven to be a unique
explanation for the anomalous flux ratios, this would confirm that the
small-scale power is high and would lend support to the view that the
problem of density distribution in galaxies has an ``astrophysical''
solution.

In order to extract useful constraints from lensing observations, we must
understand what parameters the lensing is sensitive to and what type
of halos may contribute to the total lensing signal.  In this paper,
we presented results of the first study comparing the lensing optical
depths due to small-mass halos within the lens halo (substructure), the
halos in the immediate vicinity of the lens, and the overall
cosmological population of small-mass halos distributed throughout the
entire line of sight between the source and the observer.

The main result of our paper is that the dominant contribution to
the total lensing optical depth is provided by the DM clumps in the
densest, innermost regions of the lens halo. The optical depth is
therefore very sensitive to the spatial distribution of substructure
clumps within the lens. While the effect of masses near the lens
galaxy is more important than elsewhere, the cumulative effect of all
the (isolated) halos along the line-of-sight to the source is not
negligible.  In the $\Lambda$CDM universe, we find that the
contribution of isolated halos can be a sizeable fraction of the total
lensing signal. 

The exact percentage of the isolated halo contribution depends sensitively on the mass function of
substructure halos and, especially, on their spatial distribution. The
latter is currently uncertain in cosmological simulations. In the
highest resolution cosmological $N$-body simulations reported to date,
no clumps are typically found within the central $10\%$ of the virial
radius, while at larger radii the number density profile of
substructure halos has the shape similar to that of the overall DM
profile \citep{colin_etal99,ghigna_etal00,stoehr02,kravtsov_etal03}.
If such a distribution is assumed in our model, the contribution of isolated
halos to the optical depth can be as high as $\sim 20-30\%$.  The degree to which
the results of cosmological simulations are affected by resolution is
currently unclear. It is possible that in the innermost regions the
simulations still suffer from the perennial ``overmerging'' problem.
On the other hand, the tidal disruption of NFW halos by the host is 
expected \citep{klypin_etal99b,hayashi_etal02}.  In
addition, clumps in real lenses would experience enhanced tidal forces
due to the baryonic material in the center of the lens which would
make their destruction more efficient than in $N$-body simulations, and so the lack of halos
in the central 10\% of the virial radius may be a real effect. 

Our results imply that flux ratio anomalies, if caused by DM
clumps, do indeed probe the substructure population of lens halos
and are a very promising test of CDM predictions on small-scales.
For example, the alternative scenarios proposed to remedy perceived
CDM problems, such as the simplest variants\footnote{More sophisticated
SIDM models in which interaction cross-section varies with particle
velocity may produce substructure populations similar to those of
the CDM models \citep{colin_etal02}.} of self-interacting DM
\citep[SIDM][]{spergel_steinhardt00} and warm DM \citep[WDM,
e.g.][]{colin_etal00,bode_etal01} predict a very reduced abundance of
substructure halos and, thus, a lower lensing optical depth compared to 
that from the CDM models. 

Our estimates for the optical depth in the $\Lambda$CDM halos are
in the range of $\sim 0.2-20$ and could naturally explain the
high-frequency of anomalous flux ratio images.  In contrast, SIDM and WDM
models would have optical depths of $\tau\lesssim 0.1$, although
detailed calculations are needed.  In addition in WDM and SIDM 
scenarios, lensing by
isolated halos and by substructure would have comparable optical depths because we do not
expect a significant change in the abundance of isolated halos in
these models. To be precise, WDM models predict that abundances
should be suppressed below a mass scale that corresponds to the scale
of the power spectrum cut-off. The suppression is applicable for both
isolated and substructure halos. However, for substructure halos the
abundance is further decreased by tidal disruption, while the abundance
of the field halos may be increased due to fragmentation \citep[e.g.,][]{colin_etal00,bode_etal01,knebe_etal03}.  So we expect that
the contribution of isolated halos would be even more important in the
WDM models than in $\Lambda$CDM. 

We should also note that our results imply that precise cosmological
constraints, such as estimates of the WDM particle mass or of 
the cross-section of the SIDM interaction, are difficult if not impossible to
derive at this point. As shown in our analysis, isolated field
halos can contribute a sizeable fraction to the lensing optical depth.
Also, the ingredients that are needed to estimate relative effects of
substructure and isolated halos in various models are still rather
uncertain.  Even in the relatively well studied $\Lambda$CDM
cosmology, the distribution of DM clumps in the innermost regions of
halos is not well understood. In particular, simulations 
which include the effects of baryons (a disk or stars in an elliptical galaxy)
on the tidal disruption of DM clumps are only
starting to be employed \citep[e.g.,][]{bradac_etal02}. Even for 
isolated halos the \citet{sheth_tormen99} mass function that we used
in our calculations has been tested only down to $M\sim 10^{10}h^{-1}{\ \rm
  M_{\odot}}$ \citep{reed_etal03}.  Given the importance of the
problem, these issues should and will be addressed in the next
generation of cosmological simulations. Significant progress in
determining properties of substructure halos and the mass function of
field populations in the last several years makes us optimistic that
uncertainties will be resolved in the near future.

\acknowledgements We would like to thank Neal Dalal, Daniel Holz,
Anatoly Klypin, Andrew Zentner, and James Bullock for useful discussions
and the anonymous referee for constructive comments.
This work was partially supported by the grant AST-0206216 from the
National Science Foundation (NSF).  C.R.K.\ is supported by NASA
through Hubble Fellowship grant HST-HF-01141.01-A from the Space
Telescope Science Institute, which is operated by the Association of
Universities for Research in Astronomy, Inc., under NASA contract
NAS5-26555.

\bibliography{tau}

\begin{thebibliography}{44}
\expandafter\ifx\csname natexlab\endcsname\relax\def\natexlab#1{#1}\fi

\bibitem[{{Benson} {et~al.}(2002){Benson}, {Lacey}, {Baugh}, {Cole}, \&
  {Frenk}}]{benson_etal02}
{Benson}, A.~J., {Lacey}, C.~G., {Baugh}, C.~M., {Cole}, S., \& {Frenk}, C.~S.
  2002, \mnras, 333, 156

\bibitem[{{Bergstr{\" o}m} {et~al.}(1999){Bergstr{\" o}m}, {Edsj{\" o}},
  {Gondolo}, \& {Ullio}}]{bergstrom_etal99}
{Bergstr{\" o}m}, L., {Edsj{\" o}}, J., {Gondolo}, P., \& {Ullio}, P. 1999,
  \prd, 59, 43506

\bibitem[{{Bode} {et~al.}(2001){Bode}, {Ostriker}, \& {Turok}}]{bode_etal01}
{Bode}, P., {Ostriker}, J.~P., \& {Turok}, N. 2001, \apj, 556, 93

\bibitem[{{Brada{\v c}} {et~al.}(2002){Brada{\v c}}, {Schneider}, {Steinmetz},
  {Lombardi}, {King}, \& {Porcas}}]{bradac_etal02}
{Brada{\v c}}, M., {Schneider}, P., {Steinmetz}, M., {Lombardi}, M., {King},
  L.~J., \& {Porcas}, R. 2002, \aap, 388, 373

\bibitem[{{Bullock} {et~al.}(2001){Bullock}, {Kolatt}, {Sigad}, {Somerville},
  {Kravtsov}, {Klypin}, {Primack}, \& {Dekel}}]{bullock_etal01}
{Bullock}, J.~S., {Kolatt}, T.~S., {Sigad}, Y., {Somerville}, R.~S.,
  {Kravtsov}, A.~V., {Klypin}, A.~A., {Primack}, J.~R., \& {Dekel}, A. 2001,
  \mnras, 321, 559

\bibitem[{{Bullock} {et~al.}(2000){Bullock}, {Kravtsov}, \&
  {Weinberg}}]{bullock_etal00}
{Bullock}, J.~S., {Kravtsov}, A.~V., \& {Weinberg}, D.~H. 2000, \apj, 539, 517

\bibitem[{{Col{\'{\i}}n} {et~al.}(2000){Col{\'{\i}}n}, {Avila-Reese}, \&
  {Valenzuela}}]{colin_etal00}
{Col{\'{\i}}n}, P., {Avila-Reese}, V., \& {Valenzuela}, O. 2000, \apj, 542, 622

\bibitem[{{Col{\'{\i}}n} {et~al.}(2002){Col{\'{\i}}n}, {Avila-Reese},
  {Valenzuela}, \& {Firmani}}]{colin_etal02}
{Col{\'{\i}}n}, P., {Avila-Reese}, V., {Valenzuela}, O., \& {Firmani}, C. 2002,
  \apj, 581, 777

\bibitem[{{Col{\'{\i}}n} {et~al.}(1999){Col{\'{\i}}n}, {Klypin}, {Kravtsov}, \&
  {Khokhlov}}]{colin_etal99}
{Col{\'{\i}}n}, P., {Klypin}, A.~A., {Kravtsov}, A.~V., \& {Khokhlov}, A.~M.
  1999, \apj, 523, 32

\bibitem[{{Dalal} \& {Kochanek}(2002)}]{dalal_kochanek02}
{Dalal}, N., \& {Kochanek}, C.~S. 2002, \apj, 572, 25

\bibitem[{{Eisenstein} \& {Hu}(1999)}]{eisenstein_hu99}
{Eisenstein}, D.~J., \& {Hu}, W. 1999, \apj, 511, 5

\bibitem[{Evans \& Witt(2003)}]{evans_witt03}
Evans, N.~W., \& Witt, H. 2003, \mnras, submitted (astro-ph/0212013)

\bibitem[{{Ghigna} {et~al.}(2000){Ghigna}, {Moore}, {Governato}, {Lake},
  {Quinn}, \& {Stadel}}]{ghigna_etal00}
{Ghigna}, S., {Moore}, B., {Governato}, F., {Lake}, G., {Quinn}, T., \&
  {Stadel}, J. 2000, \apj, 544, 616

\bibitem[{{Hannestad} \& {Scherrer}(2000)}]{hannestad_scherrer00}
{Hannestad}, S., \& {Scherrer}, R.~J. 2000, \prd, 62, 43522

\bibitem[{{Hayashi} {et~al.}(2002){Hayashi}, {Navarro}, {Taylor}, {Stadel}, \&
  {Quinn}}]{hayashi_etal02}
{Hayashi}, E., {Navarro}, J., {Taylor}, J., {Stadel}, J., \& {Quinn}, T. 2002,
  preprint (astro-ph/0203004)

\bibitem[{{Hu} {et~al.}(2000){Hu}, {Barkana}, \& {Gruzinov}}]{hu_etal00}
{Hu}, W., {Barkana}, R., \& {Gruzinov}, A. 2000, \prl, 85, 1158

\bibitem[{{Impey} {et~al.}(1998){Impey}, {Falco}, {Kochanek}, {Leh{\' a}r},
  {McLeod}, {Rix}, {Peng}, \& {Keeton}}]{impey_etal98}
{Impey}, C.~D., {Falco}, E.~E., {Kochanek}, C.~S., {Leh{\' a}r}, J., {McLeod},
  B.~A., {Rix}, H.-W., {Peng}, C.~Y., \& {Keeton}, C.~R. 1998, \apj, 509, 551

\bibitem[{{Keeton}(2001)}]{keeton01}
{Keeton}, C.~R. 2001, preprint (astro-ph/0111595)

\bibitem[{{Keeton}(2003)}]{keeton03}
---. 2003, \apj, 584, 664

\bibitem[{{Keeton} {et~al.}(2000){Keeton}, {Christlein}, \&
  {Zabludoff}}]{keeton_etal00}
{Keeton}, C.~R., {Christlein}, D., \& {Zabludoff}, A.~I. 2000, \apj, 545, 129

\bibitem[{{Klypin} {et~al.}(1999{\natexlab{a}}){Klypin}, {Gottl{\" o}ber},
  {Kravtsov}, \& {Khokhlov}}]{klypin_etal99b}
{Klypin}, A., {Gottl{\" o}ber}, S., {Kravtsov}, A.~V., \& {Khokhlov}, A.~M.
  1999{\natexlab{a}}, \apj, 516, 530

\bibitem[{{Klypin} {et~al.}(2001){Klypin}, {Kravtsov}, {Bullock}, \&
  {Primack}}]{klypin_etal01}
{Klypin}, A., {Kravtsov}, A.~V., {Bullock}, J.~S., \& {Primack}, J.~R. 2001,
  \apj, 554, 903

\bibitem[{{Klypin} {et~al.}(1999{\natexlab{b}}){Klypin}, {Kravtsov},
  {Valenzuela}, \& {Prada}}]{klypin_etal99}
{Klypin}, A., {Kravtsov}, A.~V., {Valenzuela}, O., \& {Prada}, F.
  1999{\natexlab{b}}, \apj, 522, 82

\bibitem[{{Knebe} {et~al.}(2003){Knebe}, {Devriendt}, {Gibson}, \&
  {Silk}}]{knebe_etal03}
{Knebe}, A., {Devriendt}, J., {Gibson}, B., \& {Silk}, J. 2003, submitted
  (astro-ph/0302443)

\bibitem[{{Kravtsov} {et~al.}(2003){Kravtsov}, {Klypin}, \&
  Gottl\"ober}]{kravtsov_etal03}
{Kravtsov}, A.~V., {Klypin}, A., \& Gottl\"ober, S. 2003, in preparation

\bibitem[{{Kundic} {et~al.}(1997{\natexlab{a}}){Kundic}, {Cohen}, {Blandford},
  \& {Lubin}}]{kundic_etal97a}
{Kundic}, T., {Cohen}, J.~G., {Blandford}, R.~D., \& {Lubin}, L.~M.
  1997{\natexlab{a}}, \aj, 114, 507

\bibitem[{{Kundic} {et~al.}(1997{\natexlab{b}}){Kundic}, {Hogg}, {Blandford},
  {Cohen}, {Lubin}, \& {Larkin}}]{kundic_etal97b}
{Kundic}, T., {Hogg}, D.~W., {Blandford}, R.~D., {Cohen}, J.~G., {Lubin},
  L.~M., \& {Larkin}, J.~E. 1997{\natexlab{b}}, \aj, 114, 2276

\bibitem[{{Mao} \& {Schneider}(1998)}]{mao_schneider98}
{Mao}, S., \& {Schneider}, P. 1998, \mnras, 295, 587

\bibitem[{{Mayer} {et~al.}(2002){Mayer}, {Moore}, {Quinn}, {Governato}, \&
  {Stadel}}]{mayer_etal02}
{Mayer}, L., {Moore}, B., {Quinn}, T., {Governato}, F., \& {Stadel}, J. 2002,
  \mnras, 336, 119

\bibitem[{{Metcalf} \& {Madau}(2001)}]{metcalf_madau01}
{Metcalf}, R.~B., \& {Madau}, P. 2001, \apj, 563, 9

\bibitem[{{Metcalf} \& {Zhao}(2002)}]{metcalf_zhao02}
{Metcalf}, R.~B., \& {Zhao}, H. 2002, \apjl, 567, L5

\bibitem[{{Moore} {et~al.}(1999){Moore}, {Ghigna}, {Governato}, {Lake},
  {Quinn}, {Stadel}, \& {Tozzi}}]{moore_etal99}
{Moore}, B., {Ghigna}, S., {Governato}, F., {Lake}, G., {Quinn}, T., {Stadel},
  J., \& {Tozzi}, P. 1999, \apjl, 524, L19

\bibitem[{{Navarro} {et~al.}(1997){Navarro}, {Frenk}, \&
  {White}}]{navarro_etal97}
{Navarro}, J.~F., {Frenk}, C.~S., \& {White}, S.~D.~M. 1997, \apj, 490, 493

\bibitem[{{Patnaik} {et~al.}(1992){Patnaik}, {Browne}, {Walsh}, {Chaffee}, \&
  {Foltz}}]{patnaik_etal92}
{Patnaik}, A.~R., {Browne}, I.~W.~A., {Walsh}, D., {Chaffee}, F.~H., \&
  {Foltz}, C.~B. 1992, \mnras, 259, 1P

\bibitem[{{Patnaik} {et~al.}(1999){Patnaik}, {Kemball}, {Porcas}, \&
  {Garrett}}]{patnaik_etal99}
{Patnaik}, A.~R., {Kemball}, A.~J., {Porcas}, R.~W., \& {Garrett}, M.~A. 1999,
  \mnras, 307, L1

\bibitem[{{Peacock} \& {Dodds}(1996)}]{peacock_dodds96}
{Peacock}, J.~A., \& {Dodds}, S.~J. 1996, \mnras, 280, L19

\bibitem[{Reed {et~al.}(2003)Reed, Gardner, Quinn, Fardal, G., \&
  Governato}]{reed_etal03}
Reed, D., Gardner, J., Quinn, T., Fardal, M., G., L., \& Governato, F. 2003,
  astro-ph/0301270

\bibitem[{{Sheth} \& {Tormen}(1999)}]{sheth_tormen99}
{Sheth}, R.~K., \& {Tormen}, G. 1999, \mnras, 308, 119

\bibitem[{{Somerville}(2002)}]{somerville02}
{Somerville}, R.~S. 2002, \apjl, 572, L23

\bibitem[{{Spergel} \& {Steinhardt}(2000)}]{spergel_steinhardt00}
{Spergel}, D.~N., \& {Steinhardt}, P.~J. 2000, \prl, 84, 3760

\bibitem[{Stoehr(2002)}]{stoehr02}
Stoehr, F. 2002, private communication

\bibitem[{{Tasitsiomi} \& {Olinto}(2002)}]{iro_olinto02}
{Tasitsiomi}, A., \& {Olinto}, A.~V. 2002, \prd, 66, 83006

\bibitem[{{Tonry}(1998)}]{tonry98}
{Tonry}, J.~L. 1998, \aj, 115, 1

\bibitem[{{Weymann} {et~al.}(1980){Weymann}, {Latham}, {Roger}, {Angel},
  {Green}, {Liebert}, {Turnshek}, {Turnshek}, \& {Tyson}}]{weymann_etal80}
{Weymann}, R.~J., {Latham}, D., {Roger}, J., {Angel}, P., {Green}, R.~F.,
  {Liebert}, J.~W., {Turnshek}, D.~A., {Turnshek}, D.~E., \& {Tyson}, J.~A.
  1980, \nat, 285, 641

\end{thebibliography}

\end{document}